\documentclass[fleqn,usenatbib]{mnras}

\usepackage{newtxtext,newtxmath}
\usepackage{multirow}
\usepackage{makecell}
\usepackage{placeins}
\usepackage{booktabs, multirow} % for borders and merged ranges
\usepackage{soul}% for underlines
\usepackage[table]{xcolor} % for cell colors
\usepackage{changepage,threeparttable} % for wide tables
\usepackage{siunitx}
\usepackage{lscape}
\usepackage{longtable}
\usepackage{diagbox}
\usepackage[T1]{fontenc}

\DeclareRobustCommand{\VAN}[3]{#2}
\let\VANthebibliography\thebibliography
\def\thebibliography{\DeclareRobustCommand{\VAN}[3]{##3}\VANthebibliography}

\usepackage{graphicx}	% Including figure files
\usepackage{amsmath}	% Advanced maths commands

\title[BCG alignment]{BCG alignment with the Locations of Cluster Members and the Large Scale Structure out to 10 R$_{200}$}

\author[]{Rory Smith$^{1}$\thanks{E-mail:rorysmith274@gmail.com},
Ho Seong Hwang,$^{2,3,4}$
Katarina Kraljic$^{5,6}$
Paula Calder\'on-Castillo,$^{1}$
Thomas M. Jackson,$^{7,8}$
\newauthor
Anna Pasquali,$^{8}$
Jihye Shin,$^{4}$
Jongwan Ko,$^{4,9}$
Jaewon Yoo,$^{10}$
Hyowon Kim,$^{9}$
and Jae-woo Kim,$^{4}$
\\
$^{1}$Departamento de Física, Universidad Técnica Federico Santa María, Avenida Vicuña Mackenna 3939, San Joaquín, Santiago de Chile
\\
$^{2}$Astronomy Program, Department of Physics and Astronomy, Seoul National University, 1 Gwanak-ro, Gwanak-gu, Seoul 08826, Republic of Korea
\\
$^{3}$SNU Astronomy Research Center, Seoul National University, 1 Gwanak-ro, Gwanak-gu, Seoul 08826, Republic of Korea
\\
$^{4}$Korea Astronomy and Space Science Institute (KASI), 776 Daedeokdae-ro, Yuseong-gu, Daejeon 34055, Korea
\\
$^{5}$Universit\'e de Strasbourg, CNRS UMR 7550, Observatoire astronomique de Strasbourg, 11 rue de l’Universit\'e, 67000 Strasbourg, France
\\
$^{6}$Aix Marseille Universit\'e, CNRS, CNES, UMR 7326, Laboratoire d'Astrophysique de Marseille, Marseille, France
\\
$^{7}$Tuparev AstroTech, 3 Sofiyski geroy Str., entr. 2 app. 28, Sofia 1612, Bulgaria
\\
$^{8}$Astronomisches Rechen-Institut, Zentrum f\"ur Astronomie der Universit\"at Heidelberg, M\"onchhofstra{\ss}e 12-14, 69120 Heidelberg, Germany
\\
$^{9}$University of Science and Technology (UST), Daejeon 34113, Korea
\\
$^{10}$Quantum Universe Center, Korea Institute for Advanced Study, 85 Hoegi-ro, Dongdaemun-gu, Seoul 02455, Korea
}

\date{Accepted to MNRAS August 2023}

\pubyear{2023}

\begin{document}
\label{firstpage}
\pagerange{\pageref{firstpage}--\pageref{lastpage}}
\maketitle

\begin{abstract}
Using a sample of $>200$ clusters, each with typically $100-200$ spectroscopically confirmed cluster members, we search for a signal of alignment between the Position Angle (PA) of the Brightest Cluster Galaxy (BCG) and the distribution of cluster members on the sky about the cluster centre out to projected distances of 3~R$_{200}$. The deep spectroscopy, combined with corrections for spectroscopic incompleteness, makes our sample ideal to determine alignment signal strengths. We also use an SDSS based skeleton of the filamentary Large Scale Structure (LSS), and measure BCG alignment with the location of the LSS skeleton segments on the sky out to projected distances of 10~R$_{200}$. The alignment signal is measured using three separate statistical measures; Rao's spacing test (U), Kuiper's V parameter (V), and the Binomial probability test (P). The significance of the BCG alignment signal with both cluster members and LSS segments is extremely high (1 in a million chance or less to be drawn randomly from a uniform distribution). We investigate a wide set of parameters that may influence the strength of the alignment signal. Clusters with more elliptical-shaped BCGs show stronger alignment with both their cluster members and LSS segments. Also, selecting clusters with closely connected filaments, or using a luminosity-weighted LSS skeleton, increases the alignment signal significantly. Alignment strength decreases with increasing projected distance. Combined, these results provide strong evidence for the growth of clusters and their BCGs by preferential feeding along the direction of the filaments in which they are embedded.
\end{abstract}

\begin{keywords}
galaxies: clusters: general -- galaxies: general -- cosmology: large-scale structure of Universe\end{keywords}

%%%%%%%%%%%%%%%%%%%%%%%%%%%%%%%%%%%%%%%%%%%%%%%%%%

%%%%%%%%%%%%%%%%% BODY OF PAPER %%%%%%%%%%%%%%%%%%

\section{Introduction}\label{sec:intro}

It has long been known that, in the nearby Universe, the shape of the spatial distribution of the cluster members tends to be preferentially aligned with the position angle of the major axis (PA) of the Brightest Cluster Galaxy \citep[BCG;][]{Sastry68}, referred to as `BCG-cluster alignment'. Since then, the advent of wide-field surveys such as the Sloan Digital Sky Survey \citep[SDSS;][]{SDSSI} has enabled the study of BCG-cluster alignment using huge statistical samples of clusters, and allowed us to study which parameters dictate the strength of the alignment. \cite{NiedersteOstholt10} used SDSS DR6 data to study several thousands of clusters, and noted a tendency for richer clusters to show stronger alignment. \cite{Huang16} used SDSS DR7 \citep{Abazajian09} data for a similarly large number of clusters, that were selected using the redMaPPer cluster finding algorithm \citep{Rykoff14}. They tested a large set of parameters such as central and satellite luminosity, central size, colour, and found that the shape of the central galaxy was an important parameter. However, these studies could not spectroscopically confirm that all of their satellite samples are truly satellites, due to the shallow depth of the SDSS spectroscopy. Using the deeper spectroscopy of the GAMA survey, \cite{Georgiou2019} found a sensitive dependence of alignment strength on galaxy colour. Similarly, \cite{Rodriguez2022} used SDSS DR16 data and found that central colour of the group was a key parameter, although central shape and group mass did not play a clear role in their results. While many studies primarily focused on the locations of satellites based on optical observations, it has since been shown that the BCG-cluster alignment is also revealed when observing the cluster shape with X-rays, the Sunyaev–Zeldovich Effect, and using gravitational lensing as well \citep{Donahue16,Yuan2022}. 

 There is some debate in the literature over whether the BCG-cluster alignment strength evolves with time. Although limited by the redshift range of their sample (typically less than $z=0.4$), \cite{NiedersteOstholt10} and \cite{Hao2011} found a significant reduction in alignment strength towards higher redshifts. However, over a similar redshift range, \cite{Huang16} failed to find any significant evolution. The \cite{West17} sample spans a much larger redshift range, from $z=0.19$ to 1.8, and they found evidence for alignment in their highest redshift clusters, when the Universe was only one third of its current age. Recently, the evolution of BCG-cluster alignment was studied using cosmological simulations, and it was found that alignment has been in place since $z \sim 4$, and there has been little evolution of its strength since $z = 2$ \citep{RagoneFigueroa20}.

The main theories for the origin of BCG-cluster alignment are: primordial alignment with the surrounding matter distribution at the time of galaxy formation, gravitational torques that gradually align galaxies with the local tidal field, and/or anisotropic infall of matter into clusters along preferred directions \citep{Catelan96,West94,Libeskind13}. In a realistic cosmological setting, these different origin scenarios cannot occur fully independently of each other \citep{Faltenbacher2008}. Thus, the observed phenomena are likely an inseparable combination of all three theories. For example, West (1994) noted that the mergers which build up BCGs do not occur haphazardly, but rather along preferred axes related to large-scale anisotropies in the primordial density field. The significance of mergers occurring along preferred directions is further emphasised in \cite{Wittman19}. Using a sample of clusters undergoing major mergers, they found that the cluster shape is aligned with the merger axis, defined by a line joining the two brightest galaxies in the cluster. Using cosmological simulations of clusters, \cite{RagoneFigueroa20} found that major mergers can either strengthen, weaken or have no effect on the BCG alignment with the cluster, depending on the direction of the accretion. However, interestingly BCGs that become misaligned by a merger tended to reorientate themselves back into alignment on several gigayear timescales, in part due to tidal torques. 

The importance of mergers in driving the BCG-cluster alignment is further emphasised by the fact that the alignment strength decreases significantly when the PA of the second brightest cluster galaxy is used instead. Indeed, the alignment disappears altogether if the third brightest cluster galaxy or fainter galaxies are used \citep[][although see \citealt{Huang18}]{Torlina2007,NiedersteOstholt10, Sifon15,West17}. This underlines the unique conditions the BCGs experience in their clusters. As they are generally the central galaxy of the cluster, they suffer many mergers as the cluster grows by accretion.

 This subsequent feeding of clusters along preferential axes means the cluster shape must have some dependence on the surrounding large-scale environment. Environmental density appears to play a role in the strength of the BCG-cluster alignment \citep{Wang18}, and simulations show that the halo shape is also a function of environmental density \citep{RagoneFigueroa07}. Using the cross-correlation of Lick galaxy counts, \cite{Argyres86} and \cite{Lambas88} note a preference for alignment with the BCG PA out to distances $\sim 15$ Mpc. Similarly, \cite{Paz2011} find a signal of correlation between cluster shape and the surrounding galaxies out to 30~Mpc. The filamentary structure of the large-scale environment provides a natural manner by which clusters may be fed preferentially down the filament axes. In simulations, \cite{Codis2018} find that in particular massive galaxies tend to have spins that are orthogonal to, and shapes that are extended along their filaments, and the strength of this coherence increases with time. This form of spin alignment is driven by spin reorientation due to mergers \citep{Welker2018}. The infall of galaxies down filaments is thought to generate an overall rotation in some observed clusters \citep{Song2018}. Simulations also show that clusters connected to larger numbers of filaments are more elliptical, later formed, and more unrelaxed \citep{Ford2019,Gouin2021}. Furthermore, galaxies with more connections have been shown (in observations and simulations) to be more elliptical, redder and with lower specific star formation rates \citep{kraljic20}. However, in the complex tidal field of super-clusters, some clusters are observed to have become disconnected from the Large Scale Structure (LSS), leaving behind orphaned filaments \citep{Einsasto2020,Einasto2021}.

In this study, we seek to directly link the position angle of cluster BCGs to their cluster shape as measured using spectroscopically classified cluster members. One advantage of our study is that our cluster sample has been the subject of several extensive spectroscopic surveys (e.g. \citealt{hecs,hecs-red,hecs-sz,hwang14}) with MMT/HectoSPEC. This provides us with a large sample ($\sim200$) of clusters, each with spectroscopically confirmed cluster members out to several R$_{200}$ from the cluster centre. In this way, we can test for a signal of alignment between BCG PA and cluster members with a large sample of clusters with a reliable and deep sample of their members. Additionally, the availability of SDSS imaging allows us to include corrections for spectroscopic incompleteness to try to correct our alignment signal measures for uneven completeness about the cluster centre.

For a selection of these clusters, we complement this data with a map of the Large Scale Structure, built using redshifts from the main galaxy sample of the SDSS survey using the code DisPerSE \citep[Discrete Persistent Structure Extractor code;][]{Sousbie11}. This allows us to trace out the filamentary skeleton of the LSS out to 10~R$_{200}$, in order to test for alignment between the BCG PA at radii far beyond the cluster vicinity, as well as consider how parameters such as the number of filament connections and distance to closest filament impact on the alignment strength.

This work is organized as follows. In Section 2, we describe our sample selection and in Section 3 we explain our method. In Section 4 we present our results. Finally, in section 5, we discuss the results and summarise them.

Throughout the paper, we adopt a standard $\Lambda$CDM cosmology with $\Omega_{\rm{M}}=0.3$, $\Omega_\Lambda=0.7$ and $h$ = 0.7. Magnitudes are given in the AB system. All the results here are based on the 16th Data Release of the Sloan Digital Sky Surveys (DR16, \citealt{SDSSdr16}).

\section{Sample}
\label{samplesection}

\subsection{The Cluster Sample}
 Our total sample consists of 211 X-ray selected clusters where we have dedicated redshift surveys 
 other than the SDSS within the SDSS footprint.
 These include 58 from the HectoSPEC Cluster Survey (HeCS; \citealp{hecs}), 121 from HeCS-red \citep{hecs-red}, 123 from HeCS-SZ \citep{hecs-sz}, 9 from the weak-lensing cluster survey \citep{hwang14}, and 2 clusters from 
 OmegaWINGS \citep{omegawings}.
 We have also included 72 Cluster Infall Region Survey clusters (CIRS; \citealp{cirs}) that have the SDSS data along with the redshifts from
 the NASA/IPAC Extragalactic Database (NED\footnote{The NASA/IPAC Extragalactic Database (NED) is operated by the Jet Propulsion Laboratory, California Institute of Technology,
under contract with the National Aeronautics and Space Administration.}).
 As there are several clusters in common between the various surveys,
 the final number of clusters is 211.
 These clusters were mainly selected to take advantage of simultaneous coverage by SDSS for optical wavelengths and ROSAT \citep{ROSAT} for X-ray coverage. The clusters range in mass from M$_{200}=1 \times 10^{13}$~M$_\odot$/h to $1 \times 10^{15}$~M$_\odot$/h (although 70$\%$ are between $1 - 5 \times 10^{14}$~M$_\odot$/h). These masses were determined dynamically using the caustic technique, while the cluster centres are determined based on their X-ray centres (see \citealp{hecs} for details). They range in redshift from $z= 0.003 - 0.289$ with a mean of $z= 0.117$. A full list of the clusters and their properties can be found in columns (i)-(vii) of Table \ref{tab:clusterlist}, and the redshift survey from which they are drawn is given in column (x).

 All the redshift estimates are collected from the literature listed above, and combined with those in the SDSS DR16. Typically, MMT/HectoSPEC spectroscopy from those papers was conducted on $\sim400-600$ candidate cluster members per cluster. These cluster member candidates were mainly selected using the red sequence technique \citep{RedSeqTech}, based on SDSS panchromatic photometry, meaning the selected galaxies are generally biased towards early-type galaxies. We also supplement these redshifts with additional redshift measurements from NED which are typically less biased toward early-type galaxies. Cluster members were then identified using the caustic technique \citep{Diaferio1999}. The number of cluster members found varies widely between individual clusters (from $\sim20$ to greater than 1000 and depends on the cluster mass), but generally our clusters have large numbers of identified members. For example, 70\% of the clusters have $>100$ spectroscopically confirmed cluster members. The wide field-of-view of HectoSPEC allows cluster members to be typically identified out to several R$_{200}$ from the clusters. 193 clusters (91\%) have confirmed members beyond 2 R$_{200}$ and 149 clusters (70\%) have confirmed members beyond 3 R$_{200}$. Targets were prioritised according to their apparent magnitude and their distance from the cluster centre. This means galaxy completeness falls rapidly for galaxies fainter than $\rm r=19.5$ for HeCS clusters (17.77 for CIRS clusters), hence we cut our sample for galaxies fainter than these limiting magnitudes. The limiting magnitude for each cluster is given in column (viii) of Table \ref{tab:clusterlist}. We further correct for incompleteness by comparing to SDSS galaxy catalogues in the same field. We can then calculate the spectroscopic completeness for the galaxies in the cluster field, which is measured in R$_{200}$-sized square apertures on the sky around the cluster's location. With this, we can introduce corrections to our counts of galaxies on the sky about the cluster centre. This reduces the impact of uneven completeness as a function of distance from the cluster centre and for varying position angle on the sky on our measured alignment signals. We also test if the strength of the alignment depends on a measure of the overall cluster completeness in Section \ref{sec:membersonlyresults}.
 
 For BCGs and all cluster members, we also consider their  position angle on the sky about the cluster centre, and their galaxy shape ($b$/$a$ axial ratio). The galaxy shape is from the best-fit parameters of the SDSS images with the de Vaucouleurs fit, which are provided by the SDSS Skyserver DR16 \citep{SDSSpara}.
 We adopt these values in the r-band as the use of a filter which collects red optical light means we are less impacted by bright star-forming regions, and the data is relatively deep. We compare to measurements in the i-band and generally find very good agreement (e.g., 90\% of PAs agree within 2 degrees). In general, the criteria for choosing the BCG within the cluster is to select the brightest spectroscopically confirmed cluster member within a projected radius of 0.5 R$_{200}$ of the cluster centre. We visually check each case as occasionally the brightest object is a star that has been misclassified as a galaxy, or the BCG was too bright to be included in the SDSS redshift survey, or the galaxy shape and position angle is poorly defined due to close galaxy pairs or pixel-bleeding. The BCG parameters for each cluster are given in columns (xi)-(xv) of Table \ref{tab:clusterlist}.

\subsection{The Large Scale Structure Surrounding Clusters}
\label{LSSsection}

In order to search for alignment between the BCG PA with the Large Scale Structure (LSS) surrounding the cluster, we first build an LSS map based on galaxy redshifts from the SDSS survey. Initially, we choose a sample of galaxies brighter than the magnitude limit of the SDSS main galaxy sample for uniformity (i.e. $m_r<17.77$; \citealt{strauss02}). 
We first run the Friends-of-Friends algorithm with a variable linking length 
 following \citet{tempel14}.
This is to take into account the effect that the galaxy number density 
  changes with redshift in the magnitude-limited sample. 
We therefore compute the mean galaxy separation at each redshift 
  ($d_{\rm mean}$), and adopt the linking lengths of 
  $0.2~d_{\rm mean}$ perpendicular to the line of sight and 
  $1~d_{\rm mean}$ along the line of sight. 
These linking lengths correspond to 1 and 5 $h^{-1}$ Mpc, respectively, at the median redshift of the sample (i.e. $z\sim0.1$). We can then correct the 3D location of these galaxies for the  finger-of-god effect by assuming that the group shape and dispersion is symmetrical perpendicularly and along our line of sight \citep[][see also \citealt{tegmark04,hwang16}]{Kraljic2018}.

Now, with the 3D locations of the galaxies, we extract the skeleton of the filamentary LSS structure surrounding our clusters using the publicly available code DisPerSE \citep{Sousbie11,SousbiePichon2011}\footnote{\href{http://www2.iap.fr/users/sousbie/web/html/indexd41d.html}{http://www2.iap.fr/users/sousbie/web/html/indexd41d.html}}, run
with a $5~\sigma$ persistence threshold on the distribution of galaxies. This choice allows us to easily identify the larger and more dense filaments typically found connected to clusters. A lower value  of the persistence would recover less substantial filaments, and potentially inject more noise into the alignment measurement.

As more massive galaxies tend to be found closer to filaments, we weight the Delaunay tessellation \citep[see][for detailed description of the DisPerSE code]{Sousbie11} by the galaxy luminosities. In Section \ref{sec:LSSsubsamples}, we will see that weighting by galaxy luminosity can significantly influence the strength of the alignment signal.

Although we have information on the cluster members for the full sample of 211 clusters, our LSS skeleton is only constructed for objects within the main SDSS area and some of the HeCS clusters fall outside this footprint. Those clusters that fall inside this footprint are labelled `LSS' in column (xvi) of Table \ref{tab:clusterlist}. Therefore, we cannot study the alignment with the LSS for all of the clusters. Also, we require that the LSS skeleton connects directly with the cluster in three-dimensions -- our criteria is that at least one segment must cross a sphere of 3 R$_{200}$ centred on each cluster (R$_{\rm{200}}$,connect = 3~R$_{200}$), and we exclude any unconnected filaments from our analysis. In Section \ref{sec:LSSsubsamples}, we test the sensitivity of our results to this fairly arbitrary choice of value for R$_{\rm{200}}$,connect. Finally, when studying the BCG PA alignment with the LSS, we exclude clusters that have another cluster within 5 R$_{200}$ (Near Pair, NP$>$5 R$_{200}$). This reduces the presence of numerous galaxies or LSS segments associated with the other cluster as a noise source in our measured alignment signal. Additionally, we test how this choice of distance affects our results in Section \ref{sec:LSSsubsamples}. 

\subsection{The `Cluster Members-only' sample versus the `Cluster Members + LSS' sample}
\label{sec:samples}

As a result of the various restrictions described in Section \ref{LSSsection}, our full sample of 211 clusters with spectroscopically confirmed members is reduced to 91 clusters, if we require the information on their surrounding LSS for our analysis. Thus, we refer to this sample as the `Cluster Members + LSS' sample, and we always use this sample when studying BCG alignment with the LSS, and the dependency of the alignment on LSS parameters (e.g., number of filament connections, etc). We consider the results for BCG alignment measured using this sample in Section \ref{sec:visualLSSresults} and \ref{sec:LSSsubsamples}.

However, we can use the full sample of 211 clusters if we restrict ourselves to consider alignment between the BCG PA and their cluster members only, neglecting the alignment with the LSS. We refer to this sample as the `Cluster Members-only' sample, and consider this sample's alignment results separately in Section \ref{sec:membersonlyresults}. By using the `Cluster Members-only' sample, we can greatly increase our number statistics. For example, the fiducial model of the `Cluster Members-only' sample consists of 13,741 galaxies compared to 5,917 member galaxies in the `Cluster Members + LSS' sample.

\section{Method}
\label{sec:method}
We can now begin to search for a signature of alignment between the PA and the location of cluster members/LSS segments on the sky. Our method for doing this is as follows. 

We first measure the typical strength of the alignment for the cluster sample as a whole. This is accomplished by stacking multiple clusters together, after rotating each cluster individually so as their BCG PAs are aligned vertically on the stacked image. By stacking, we can average out some of the noise that may exist in individual clusters, and we can also greatly increase the number of cluster members and/or LSS segments that are used to measure the strength of the alignment. 

Then, we can simply compute each satellite galaxy's PA on the sky, measured from the cluster centre. We then calculate the difference between this PA and that of the BCG PA. 

For the LSS, the approach is similar. 
DisPerSE filaments consist of a series of linear segments with two endpoints. To measure the BCG alignment with the segments, we measure the position angle of each segment on the sky, measured with respect to the cluster centre, and compare with the BCG PA. We note that an alternative means to measure the alignment with the segments would be to measure the position angle of one end of a segment from the other end, and compare that with the BCG PA. However, this alternative approach is likely more sensitive to segment-to-segment deviations\footnote{Testing with this alternative measure recovers qualitatively similar results.}. Therefore, in this study we focus on the use of segment positions on the sky about the cluster centre instead.

We wish to know if the alignment signal extends to distances far beyond the influence of the cluster. Therefore, we divide our sample into radial bins by their projected radius from the cluster centre. For the LSS segments the radial bins are $0-3$~R$_{200}$, $3-6$~R$_{200}$ and $6-10$~R$_{200}$. For the cluster members, we test radial bins of $0-1$~R$_{200}$, $1-2$~R$_{200}$ and $2-3$ R$_{200}$. We also consider the case where all the radial bins are combined ($0-10$ R$_{200}$ for the LSS segments, and $0-3$ R$_{200}$ for the cluster members) in order to maximise the statistics.

Following \cite{West17}, we consider three separate statistical methods to quantify the strength of the BCG alignment signal; the Rao-spacing test, Kuiper's V-statistic and the Binomial test (referred to as the measures $U$, $V$ and $P$ respectively from herein). One strength of these methods is that there is no requirement for arbitrary binning of the PAs of the galaxies/LSS segments about the cluster center.

\subsection{The Rao-Spacing test, U}
\label{sec:Umeas}
For a sample of size $N$ with a perfectly uniform distribution, galaxies/LSS segments should have offsets from the BCG PA that are evenly spaced between 0 and 90 degrees: 

\begin{equation}
    \lambda=\frac{90}{N}\textrm{,}
\end{equation}
\noindent
where $\lambda$ is the expected spacing of a uniform distribution. The Rao-spacing test measures the amount of deviation from the case of even spacing, and can be defined as 

\begin{equation}
    U = \frac{1}{2} \sum^{N-1}_{i=1} |T_i - \lambda|\textrm{,}
\end{equation}
\noindent
where $\theta_i$ is the angular offset of the $i$th object from the BCG PA, $T_i$=$\theta_{i+1}-\theta_i$ for $i \leq N-1$, and $T_i$=$(90-\theta_N)+\theta_1$ for $i = N$. 

In practice, the sample's PA values are sorted by size such that the angle between successive values can be easily measured. Because of the summation of deviations from the uniform case, the $U$ value is larger for a less uniform distribution. However, we note that the $U$ measure may also be enhanced if the distribution of PAs is clumpy, rather than simply preferring a single direction on the sky.

\subsection{Kuiper's V-statistic, V}
To measure Kuiper's V-statistic, the PAs are once again sorted into increasing order and a cumulative distribution of their values is made from 0 to 90 degrees. The maximum value of the cumulative distribution above that of a pure uniform distribution is assigned to the $D_{+}$ variable. The minimum value of the cumulative distribution below that of a pure uniform distribution is assigned to the $D_{-}$ variable. Kuiper's V-statistic is then simply given by their summation:

\begin{equation}
    V = D_{+} + D_{-}\textrm{.}
\end{equation}

As a result, if the galaxy/LSS segment PAs prefer a particular direction, such as the BCG PA, the value of $V$ will increase. 

\subsection{The Binomial test, P}
For the Binomial test value $P$, the fraction of objects which have a difference in PA with respect to the BCG PA of less than 45 degrees is measured. For a perfectly uniform sample, $P$=0.5, whereas if the positions of the objects on the sky show a preference for alignment with the BCG PA, $P>0.5$.

\subsection{Significance of alignment measurements}
\label{sec:significance}
Again, following the procedure outlined in \cite{West17}, we test the significance of the three alignment measures ($U$, $V$ and $P$) by calculating their probability for such a value to be drawn by chance from a uniform distribution ($P_{\rm{uniform}}$). First, we generate one million samples, of equal size to the observed sample, whose PAs are randomly drawn from a uniform distribution. For each sample, we measure its $U$, $V$ and $P$ value. Then, to calculate $P_{\rm{uniform}}$, we measure what fraction of the million samples reaches the observed values. If none of the samples reaches the observed value, the $P_{\rm{uniform}}$ must be less than one in a million (i.e., highly non-uniform). For the figures presented in this paper (Section \ref{sec:results}), we only keep data points where there is a 1 in 44 chance or less that the sample could be drawn by chance from a uniform sample (equivalent to a two-sigma or higher detection of non-uniformity). This establishes a minimum level of non-uniformity within our results, and helps to remove false trends that could be produced by chance, in particular when statistics are limited. In addition, the error bars on data points (see next section for a description) can also help us judge the significance of any observed trends.

\subsection{Comparing Subsamples of equal size}
\label{sec:equalsampsizes}
Rather than simply stacking all the clusters together, we can also experiment with subsampling the cluster sample. In this way, we can choose which clusters to include in the stack, and test dependencies of the BCG alignment signal strength on various parameters of our choice. For example, comparing high- and low-mass clusters, or producing a low-redshift cluster sample. An individual galaxy may belong to multiple subsamples (e.g., Nearby Universe and Luminosity). Also, depending on how the subsample is divided, an individual galaxy may appear on both sides of the divide (e.g., a bright galaxy in the Luminosity subsample could appear in the `All' and `Bright' category). In the same way, clusters may also belong to multiple subsamples, or both sides of the divide of a subsample, depending on how the divide is defined (e.g., a cluster will only fall in either the round or the elliptical category but it will always be found in the Fiducial category.) A full list of the subsamples considered is given in Section \ref{sec:LSSsubsamples} for the alignment with the LSS and in Section \ref{sec:membersonlyresults} for the alignment with the cluster members. 

In order to allow a fair comparison of $U$, $V$ and $P$ values between subsamples, we ensure that we only compare subsamples of equal size. The reduction in the size of a subsample is achieved by randomly selecting from the original subsample (of size $N_{\rm{orig}}$) until it reaches a sample size that is the minimum among the compared subsamples ($N_{\rm{min}}$). However, the final measurement of $U$, $V$ and $P$ may depend on which objects were selected for the reduced subsample. For the $V$ and $P$ measurements, we measure the uncertainties introduced by the subsample reduction by conducting 1000 bootstraps of the selection of $N_{\rm{min}}$ objects from the original $N_{\rm{orig}}$ objects, and measuring the mean value of $V$ and $P$ with their error given by the standard deviation. However, it is impossible to treat the $U$ values similarly, as bootstrapping results in objects in the reduced sample with repeated (identical) PA values. As $U$ is influenced by the separation between consecutive PA values that have been ordered by size (see Section \ref{sec:Umeas}), it is strongly enhanced by repeated values. Therefore, in the case of the $U$ measurement, we instead randomly select $N_{\rm{min}}$ objects from the original $N_{\rm{orig}}$ objects without any repetition of PA values. This process is repeated 1000 times and $U$ is assigned the mean value with an error given by the standard deviation. The number of objects in each of the subsamples (e.g., galaxies, LSS segments and clusters) is provided in Appendix \ref{sec:Nsamp} (Tables \ref{tab:samplesize_LSS} -- \ref{tab:Ncluster_gals}).

\section{Results}
\label{sec:results}

\subsection{Visual representation of BCG Alignment in the `Cluster Members + LSS' sample}
\label{sec:visualLSSresults}

We start by considering our sample of clusters with information on their surrounding LSS (see Section \ref{LSSsection} for details). The total number of clusters is 91.

\begin{figure*}%[ht]
\includegraphics[width=\textwidth]{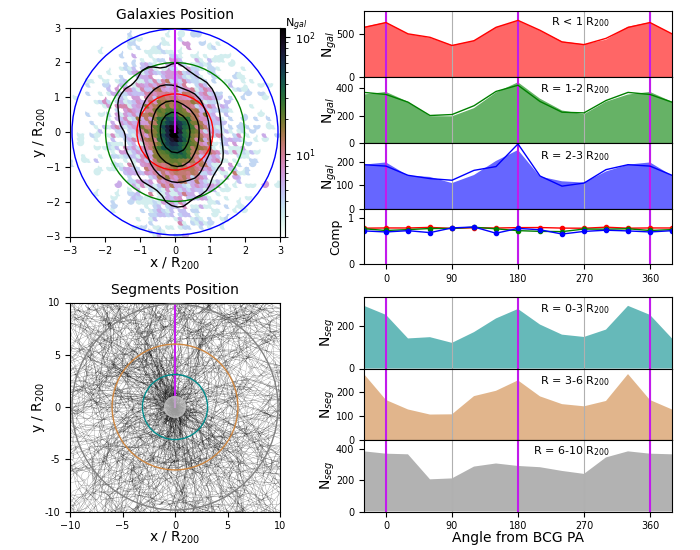}
\caption{Results of stacking all the `Cluster Members + LSS' sample. Top-left: Stack of the position of the cluster members after rotating each cluster to align their BCG PAs vertically along the purple line. The x- and y-axis are normalised by the R$_{200}$ of each individual cluster. The colour bar indicate the number of galaxies in a pixel of length 120~pc, and black iso-number contours are overlayed, to highlight the extended shape of the cluster members along the BCG PA. The coloured circles show the three bins of projected radius we consider (red, green, blue is $\rm 0-1, 1-2, 2-3~ R_{200}$, respectively). Top-right: Histograms of the angle of cluster members about the cluster centre, measured with respect to the BCG PA. The top three subpanels match the radial bins in the top-left panel. Purple vertical lines highlight where an alignment signal should result in peaks, with grey vertical lines for troughs. The lowest subpanel shows the mean completeness in each angular bin, with a different colour line for each radial bin. The shaded histograms are the renormalised completeness corrected number counts while the solid lines are not completeness corrected. Bottom-left: As in the top-left panel, but stacking the LSS segments that are connected to the cluster. We place a 1~R$_{200}$ radius translucent sphere over the cluster to try to highlight where the filaments leave from the cluster. Bottom-right: As in the top-right panel, but histograms of the number of LSS skeleton segments in each angular bin. There is no completeness subpanel as the SDSS completeness is very uniform.}
\label{fig:stackedall3Rvirconnect}
\end{figure*}

In the top-left panel of Figure \ref{fig:stackedall3Rvirconnect}, we plot the number density of the stacked cluster members (after each cluster has been rotated such that their BCG PA is always vertical). The x- and y-axis are normalised by R$_{200}$ and we overlay circles with radii of 1, 2 and 3 R$_{200}$ to better highlight the shape of the cluster member distribution which is extended along the BCG PA in both directions, as further highlighted by the contours of equal galaxy number density. In any individual cluster, the distribution of galaxies might be  more extended in one direction than in the other. However, when we stack multiple clusters, the resulting cluster member distribution tends to be symmetrical vertically. This is because, when the BCG PA is defined, it could point in either direction with equal probability.

In the top-right panel, we plot the number of cluster members in bins of position angle for all the stacked clusters. For this visual representation, we use a fairly arbitrarily chosen bin width of 30 degrees, where the first bin is positioned symmetrically about the vertical axis in the top-left panel. We note that this binning is only used for this visual demonstration of the alignment. All measures of the BCG alignment signal strength are conducted using the bin-free statistical measures that we previously described in Section \ref{sec:method}. Here, we show a separate histogram of the number of cluster members for the three projected radial bins. The elliptical shape of the cluster members distribution about the cluster centre that was visible in the upper-left panel is also clearly visible in these histograms with similar-sized peaks at 0/360 and 180 degrees.

%Therefore, the alignment signal in the cluster members drops significantly inside of 1~R$_{200}$, perhaps as a result of infalling galaxies relaxing into a more spherical distribution near the cluster centre. 

The bottom sub-panel shows that the average completeness is roughly equal in all angular bins, and thus the alignment with the satellite members is not a result of preferentially higher completeness in angle bins near the BCG PA. We present the completeness-corrected histogram as a filled area, and the histograms without completeness correction as a solid line. Note that the curve of the uncorrected histogram can be higher in some angle bins than the corrected histogram as the latter has been re-normalized to match the total sample size of the uncorrected histogram.

In the lower-left panel, we consider the stacked plot of all of the LSS skeletons that are connected to clusters at 3~R$_{200}$ either directly or indirectly. Visibly, it can be seen that there is a tendency for more filaments to emerge from the top and bottom of the clusters (i.e. parallel to the BCG PA) compared to from left to right (perpendicular to the BCG PA). We place a translucent circle of radius 1~R$_{200}$ at the centre to try to make this more visible. While the filaments that emerge from the bottom are quite vertically aligned, those emerging from the top appear slightly tilted with respect to the BCG PA. The histograms in the lower-right panel confirm that this is the case, with a well-aligned peak at 0/360 degrees in the first and second radial bin, but the other peak has a slight offset from 180 degrees. We hypothesise that this may be the result of the low number statistics of clusters considered in this stack. Nevertheless, it is clear that, far from being uniformly distributed, there is a tendency for the filaments to be more aligned with the BCG PA over the full $\rm 0-10~R_{200}$ range.

The measured $U$, $V$, and $P$ values for the stack of all the LSS segments of our `Fiducial' model (91 clusters in the sample) are $U$=35.4, $V$=684.6, and $P$=0.58 for all the radial bins combined (0-10~R$_{\rm{vir}}$). All the results are highly significant -- a by chance selection of the PA distribution from a uniform distribution is ruled out at a one in a million (or greater than  million) level (non-uniform at the level of $>$4.8$\sigma$). The alignment signal is also highly significant in the individual radial bins as well. All are non-uniform at a one in $>$one million level, except for the first radial bin of the $U$ measure which is non-uniform at a one in $>$100 thousand level (still highly significant, $>$3.9$\sigma$).

\subsection{Results of subsampling the `Cluster Members + LSS' sample}
\label{sec:LSSsubsamples}
We split the `Cluster Members + LSS' sampling according to five different parameters described below. As noted in Section \ref{sec:method}, each subsample is of equal size to enable a fair comparison of the measured U, V and P parameters, free from the effects of changing subsample size. The subsampling approach is described in Section \ref{sec:equalsampsizes}. However, this means that comparisons of alignment signal strength should only be done \textit{within a subsample, and for matching radial bins}. The results are shown in Figure \ref{fig:UVPsatsandLSS}. Data points whose probability to be uniform is too high (i.e., $P_{\rm{uniform}}$ is greater than our chosen criteria, described in Section \ref{sec:significance}) are removed from the panels. Error bars arising from the subsampling procedure are calculated as described in Section \ref{sec:equalsampsizes}.

\begin{itemize}%[leftmargin=5.5mm]

\item {\it{BCG Shape: }} The BCG shape is quantified by the axial ratio ($b$/$a$) of the BCG from the best-fit parameters of the SDSS DR16 \citep{SDSSdr16}. We split the sample into `Elliptical' BCGs (with $b$/$a < 0.75$) and `Round' BCGs ($b$/$a \ge 0.75$) in the first column of Figure \ref{fig:UVPsatsandLSS} (cluster sample sizes of 48 and 43 respectively). In the `Fiducial' subsample, no shape limit was imposed. For all three measurements ($U$, $V$ and $P$) where the radial bins are combined (0-10 R$_{200}$), the clusters with `elliptical' BCGs have a significantly higher alignment signal. For the $V$ and $P$ measures of the combined radial bins, the clusters with `round´ BCGs are those with the lowest signal strength and they are significantly lower. For example, comparing clusters with round versus elliptical BCGs, $V=180.0\pm23.0$ versus $V=439.7\pm0.27.9$, and $P=0.528\pm0.008$ versus $P=0.620\pm0.08$ respectively (all measures have $P_{\rm{uniform}}=$ 1:million or less except the $P$ measure for round BCGs which has $P_{\rm{uniform}}=$ 1:650). It is also noticeable that there is a general trend for the same dependency on signal strength to be visible in all three individual radial bins (i.e., compare matching filled symbols within the BCG shape subsamples). The BCG shape is an interesting parameter as more intrinsically elliptical BCGs could form via mergers occurring preferentially from particular directions. Given that the BCG PA shows preferential alignment with the LSS out to many R$_{200}$ from the clusters, this could be interpreted as strong evidence that filaments feed in galaxies to the cluster along their lengths and, in the process, the BCG grows along that direction. We note that if an intrinsically elliptical BCG aligns with the surrounding LSS, projection effects could cause the BCG to appear round on the sky. But this would also cause the LSS to be projected down our line of sight, and thus weaken the coherence signal. Thus, the dependency we see on BCG shape could be partly a result of projection effects, rather than the intrinsic shape of the BCG. Either way, there would be the requirement that there is some genuine alignment in three dimensions between the BCG and the LSS.  

\begin{figure*}%[ht]
\begin{center}
\includegraphics[width=165mm]{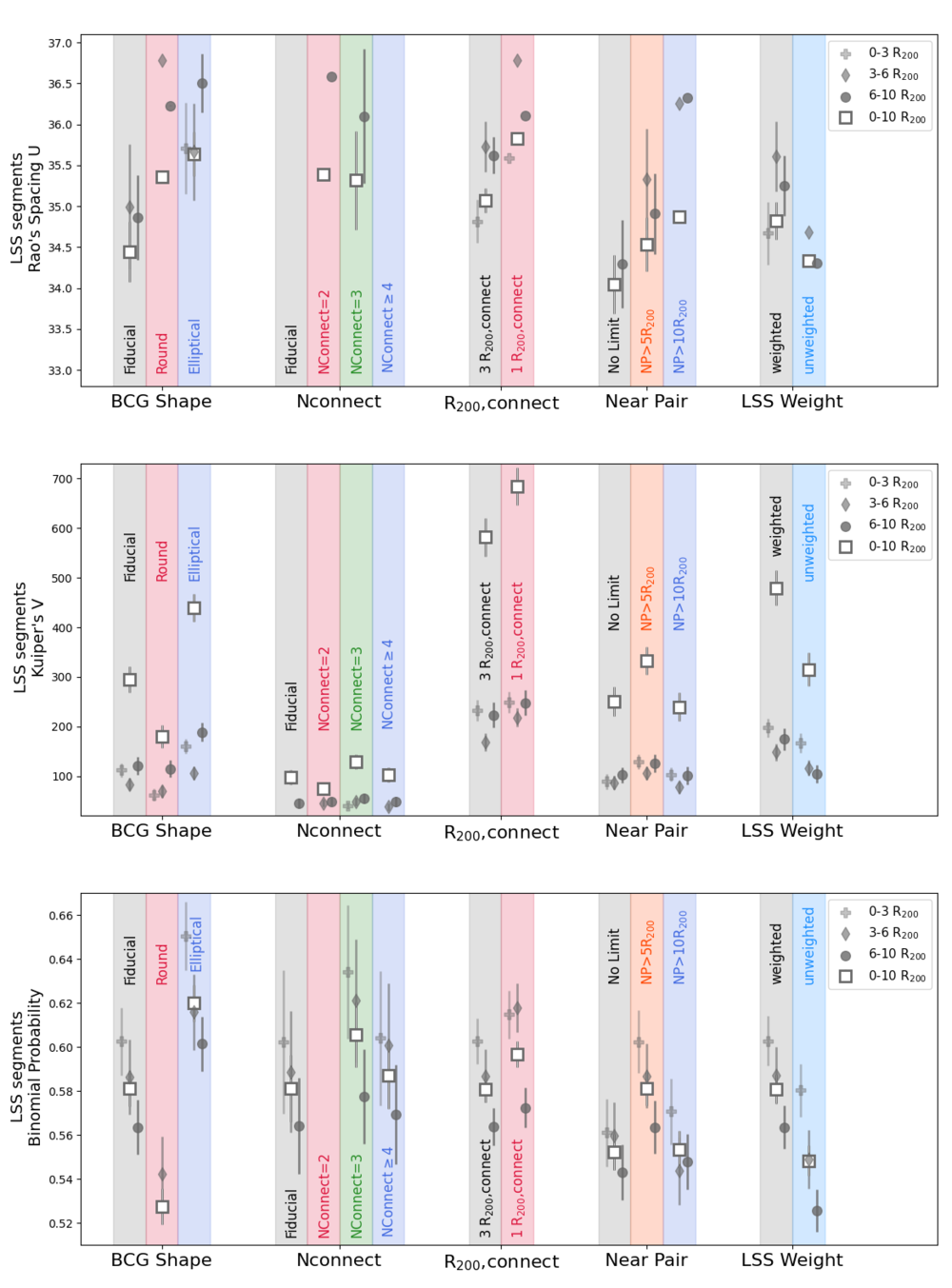}%LSS_sats_curves.png}
\caption{For the `Cluster Members + LSS' sample, comparison of the strength of the alignment signal using equal sample sizes. Symbols indicate the radial bins (see legend; $\rm 0-3~R_{200}$,  $\rm 3-6~R_{200}$, $\rm 6-10~R_{200}$, and $\rm 0-10~R_{200}$ for all the LSS segments combined). Subsamples (e.g., BCG shape) that can be compared at fixed sample size are grouped together along the x-axis. See text in Section \ref{sec:LSSsubsamples} for details on each parameter. Please note that sample sizes vary between radial bins, and between subsamples (e.g., between `BCG shape' and `Nconnect'). Therefore, comparisons should only be done within a subsample and for the same radial bin. }
\label{fig:UVPsatsandLSS}
\end{center}
\end{figure*}

\item {\it{Nconnect: }} Nconnect is the number of connecting filaments at 3~R$_{200}$ from the cluster centre. We split our sample into clusters with two, three and four or more connecting filaments (cluster sample sizes of 24, 28 and 39 respectively) in the second column of Figure \ref{fig:UVPsatsandLSS}. In the `Fiducial' subsample, no limit on numbers of connecting filaments was imposed. In general we do not see a clear dependency, although there is a hint that Nconnect=3 results in a stronger alignment signal, although it is only clearly significant for the $V$ measure when all the radial bins are combined (0-10 R$_{200}$). The lack of a strong dependency on Nconnect is interesting. Naively, we might have expected that Nconnect = 2 clusters would be better aligned with their filaments as feeding of galaxies would only be from a single filament passing through the cluster and so occur from less different directions. However we note that, by eye in the LSS maps of individual clusters, even when there are several connections (Nconnect = 3 or Nconnect $>$ 4) to a cluster, they still tend to prefer to connect with the cluster at a similar position-angle. Therefore, feeding of galaxies may still occur from preferential PAs even when there are multiple filaments connected to a cluster.

\item {\it{R$_{\rm{200}}$,connect:}} This parameter controls the radius at which the filaments connect to the cluster. In our fiducial set-up, we chose it to be at 3 R$_{200}$ from the cluster. Here, we can see the impact on the alignment signal strength if we choose it to instead be at 1~R$_{\rm{200}}$ from the cluster. For the combined radial bins (0-10 R$_{200}$), we see a significant increase in the alignment signal strength for all three measurements ($U$, $V$ and $P$) if we restrict our cluster sample to clusters with R$_{\rm{200}}$,connect=1~R$_{\rm{200}}$. This stricter criterion reduces the number of clusters in our sample from 91 to 64, but nevertheless strengthens the alignment signal. For example, comparing clusters with R$_{\rm{200}}$,connect=3~R$_{\rm{200}}$ versus R$_{\rm{200}}$,connect=1~R$_{\rm{200}}$, $U=35.07$ versus $U=35.83$ (typical errors $\sim0.15$), $V=581.7\pm$38.9 versus $V=684.3\pm38.4$, and $P=0.581\pm0.006$ versus $P=0.597\pm0.006$ respectively (where all measurements have $P_{\rm{uniform}}=$ 1:million or less). This result is generally confirmed in the individual radial bins, although with a lower significance than when all the radial bins are combined.

\item {\it{Near Pair:}} In the `Fiducial' model, we exclude clusters with a nearby pair cluster within 5 R$_{200}$, which gives a sample of 91 clusters. Here, we consider the impact of removing this limit (subsample labelled `No Limit', sample size of 124 clusters) or making the criteria much stricter by excluding clusters with a nearby pair within 10~R$_{200}$ (sample size of 55 clusters). We note that we use all 211 clusters to identify neighbour clusters. For the combined radial bins, we see that removing the limit reduces signal strength for all three signal strength measures ($U$, $V$ and $P$). For both the $V$ and $P$ measures, we see that setting the limit to 5 R$_{200}$ results in the alignment signal being the strongest and significantly so. 

\item {\it{LSS Weight:}} We normally use an LSS skeleton obtained from the galaxy distribution and the galaxy luminosity-weighted Delaunay tesselation. Here, we test the effect on the alignment signal of using an LSS skeleton that only depends on galaxy number density (labelled `unweighted' as it is not luminosity weighted). It is noticeable that the alignment signal strength for the combined radial bins (0-10 R$_{200}$) is significantly increased for all three measures ($U$, $V$ and $P$) when luminosity-weighting is used. For example, comparing the non-weighted LSS skeleton versus the weighted LSS skeleton, we measure $U=34.34$ versus $U=34.82$ (typical errors $\sim0.23$), $V=314.9 \pm 33.5$ versus $V=479.5 \pm 34.9$, and $P=0.548 \pm 0.007$ versus $P=0.581 \pm 0.007$ respectively (where all measurements have $P_{\rm{uniform}}=$ 1:million or less except the non-weighted LSS measure of $U$ which has $P_{\rm{uniform}}$= 1:110 thousand). In general, this result is also seen for the individual radial bins as well, although sometimes with less significance.

\item \textit{Dependence on Projected Distance from Cluster: }
We note that it is not fair to compare alignment signal strength measurements between radial bins of the above subsamples as, radially, we did not match them in sample size to avoid excessively reducing the number statistics. However, as an additional test, for the Fiducial model only, we match the radial bin sample sizes of LSS segments so as we can see how the signal strength varies with radius, independent of the effects of the changing galaxy numbers with radius. This reduces our statististics significantly, and, for the $U$ measure the results are not significant. But for the $V$ and $P$ measures, the results are significant. We find that the signal strength decreases significantly with increasing projected radius from the cluster for both the $V$ and $P$ measures. For the radial bins (0-3, 3-6, 6-10)R$_{200}$, $V$ = (231.8, 200.4, 158.0) with typical errors of $\sim$20.0, and $P$ = (0.603, 0.567, 0.564) with typical errors of $\sim$0.01. This suggests that the BCG PA is more strongly aligned with the nearby LSS.

\item \textit{Alignment of cluster members versus LSS segments: }
We also run an additional test that is not directly shown in Figure \ref{fig:UVPsatsandLSS}, comparing the S/N of all the cluster members combined ($\rm R < 3~R_{200}$) with the S/N of all the LSS segments combined ($\rm R<10~R_{200}$). This test is conducted only for galaxies that are members of clusters in the `Cluster Members + LSS' sample. We find that the alignment signal is stronger in the cluster member population than for the LSS segments. For example, $V$=797.8 ($P$=0.596) for cluster members versus $V$=685.97 ($P$=0.581) for the LSS segments. This result may be consistent with stronger alignment for at shorter projected distances, as the satellite population is both closer and presents a stronger alignment signal. However, it may also be partly because the original selection of our cluster members sample is biased towards redder galaxies which are known to present stronger signals of alignment (for further discussion, see Appendix \ref{sec:completeness}).

\end{itemize}

In summary, for the strength of the signal of alignment between the BCG PA and the positions of LSS segments about the cluster center, the shape of the BCG is a key parameter. Forcing the cluster to be directly connected to the LSS at 1 $R_{200}$ and using a luminosity-weighted LSS skeleton both resulted in a significant increase in alignment signal strengths.

%\begin{figure*}[ht]
%\includegraphics[width=\textwidth]{Fig_StoN_Gals_LSS.png}%S2Ncomp_LSS_sats.png}
%\caption{For the `Cluster Members + LSS' sample, comparison of the signal-to-noise of the alignment signal using equal sample sizes (denoted as S/N$_{\rm{equal}}$). The y-axis of the upper- and lower-row is the S/N$_{\rm{equal}}$ of the cluster members and the LSS segments, respectively. Symbols indicate the radial bins (see legend; $\rm 0-1~R_{200}$,  $\rm 1-2~R_{200}$, and $\rm 2-3~R_{200}$ for Cluster Members, and $\rm 0-3~R_{200}$,  $\rm 3-6~R_{200}$, and $\rm 6-10~R_{200}$ for LSS segments). Subsamples (e.g., BCG shape) that can be compared at fixed sample size are grouped together along the x-axis. See text in Section \ref{sec:LSSsubsamples} for details on each parameter. Please note that sample sizes vary between radial bins, and between subsamples (e.g., between `BCG shape' and `Nconnect'). Therefore, comparisons should only be done within a subsample and for the same radial bin.}
%\label{fig:s2Ncomparison_satsandLSS}
%\end{figure*}

\subsection{Results of subsampling the `Cluster Members only' sample}
\label{sec:membersonlyresults}
We first measure the alignment signal for all of the cluster members combined into a single stack of all 211 clusters. For all the radial bins combined (0-3~R$_{\rm{vir}}$), we measure $U$=33.6, $V$=1767.5, and $P$=0.57. The $V$ and $P$ measures are non-uniform at a one in $>$one million level (very high significance, $>$4.8$\sigma$) and the same is true for all the individual radial bins (0-1~R$_{\rm{vir}}$, 1-2~R$_{\rm{vir}}$, and 2-3~R$_{\rm{vir}}$). The $U$ measure is less significant compared to the $V$ and $P$ measure. For all radial bins combined, it is non-uniform at a 1:1350 level (3.2$\sigma$). For the individual radial bins 0-1~R$_{\rm{vir}}$, 1-2~R$_{\rm{vir}}$ and 2-3~R$_{\rm{vir}}$, it is non-uniform at the level 1:400 (2.8$\sigma$), 1:25 (1.8$\sigma$) and 1:70 (2.2$\sigma$) respectively.

Now, we split the `Cluster Members only' sample into subsamples according to the seven different criteria described below. Some of these criteria match those used in Figure \ref{fig:UVPsatsandLSS} (e.g., BCG Shape, Near Pair). But, we now consider some new subsamples that are unrelated to the LSS (e.g., Cluster Mass, member Luminosity, etc).

Once again for clarity, we note that each subsample is of equal size to enable a fair comparison of the measured $U$, $V$ and $P$ measures, free from the effects of changing subsample size. However, this means that comparisons of alignment signal strength should only be done \textit{within a subsample, and for matching radial bins}. We neglect data points representing samples whose probability to be uniform is too high
(i.e., $P_{\rm{uniform}}$, as described in Section \ref{sec:significance}), and error bars arising from the subsampling procedure are calculated as described in Section \ref{sec:equalsampsizes}. The results are shown in Figure \ref{fig:UVPsatsonly}. Our main sample (labelled `All' in the legends), consists of the full 211 clusters, and more than 13,000 spectroscopically confirmed cluster members.

%\noindent

\begin{itemize}%[leftmargin=*]

\item {\it{BCG shape: }} The BCG shape is classified in the same way as in Section \ref{sec:LSSsubsamples}, creating a sample of 106 (105) elliptical (round) BCGs. As we saw in Section \ref{sec:LSSsubsamples}, there appears to be a visibly stronger alignment signal when we consider clusters with `elliptical' BCGs compared to `round' BCGs (see first column of Figure \ref{fig:UVPsatsonly}). However, in this case, the alignment is with the cluster members rather than with LSS segments. For satellites in all the radial bins combined (0-3 R$_{200}$), this is visible for all three measures ($U$, $V$ and $P$), and is highly significant for the $V$ and $P$ measures. For example, comparing the clusters with round versus elliptical BCGs, we measure $U=33.33$ versus $U=34.33$ (typical errors $\sim0.15$), $V=510.2 \pm 50.5$ versus $V=1153.3 \pm 53.3$, and $P=0.541 \pm 0.005$ versus $P=0.597 \pm 0.005$ respectively (where all measurements have $P_{\rm{uniform}}=$ 1:million or less except the $U$ measure for round BCGs which has $P_{\rm{uniform}}$ of only 1:9 and so is excluded from the plot). Similar results are seen in all radial bins, albeit with lowered significance.

%\noindent
\item {\it{Cluster Mass: }}
In the second column of Figure \ref{fig:UVPsatsonly}, we divide the main sample into a high- and low-mass cluster sample (divided below a cluster mass of $\rm 3 \times 10^{14}~M_\odot$). The cluster sample size is 137 (74) low (high)-mass clusters. Although there are many more low-mass clusters, they also contain less cluster members, therefore the galaxy subsample sizes are more similar in size. We do not see evidence for a significant dependence on cluster mass in these panels in either the combined radial bins or individual radial bins.

%\noindent
\item {\it{Cluster Member Luminosity: }}
 In the third column of Figure \ref{fig:UVPsatsonly}, we split the sample into bright (M$_{\rm{r}}<-20.5$) and faint (M$_{\rm{r}}>-20.5$) cluster member subsamples, using the full 211 clusters. For the combined radial bins only (0-3 R$_{200}$), there is a hint that more luminous galaxies show slightly stronger alignment signal for $U$, $V$ and $P$ measures but it is of low significance.

\begin{figure*}%[ht]
\begin{center}
\includegraphics[width=165mm]{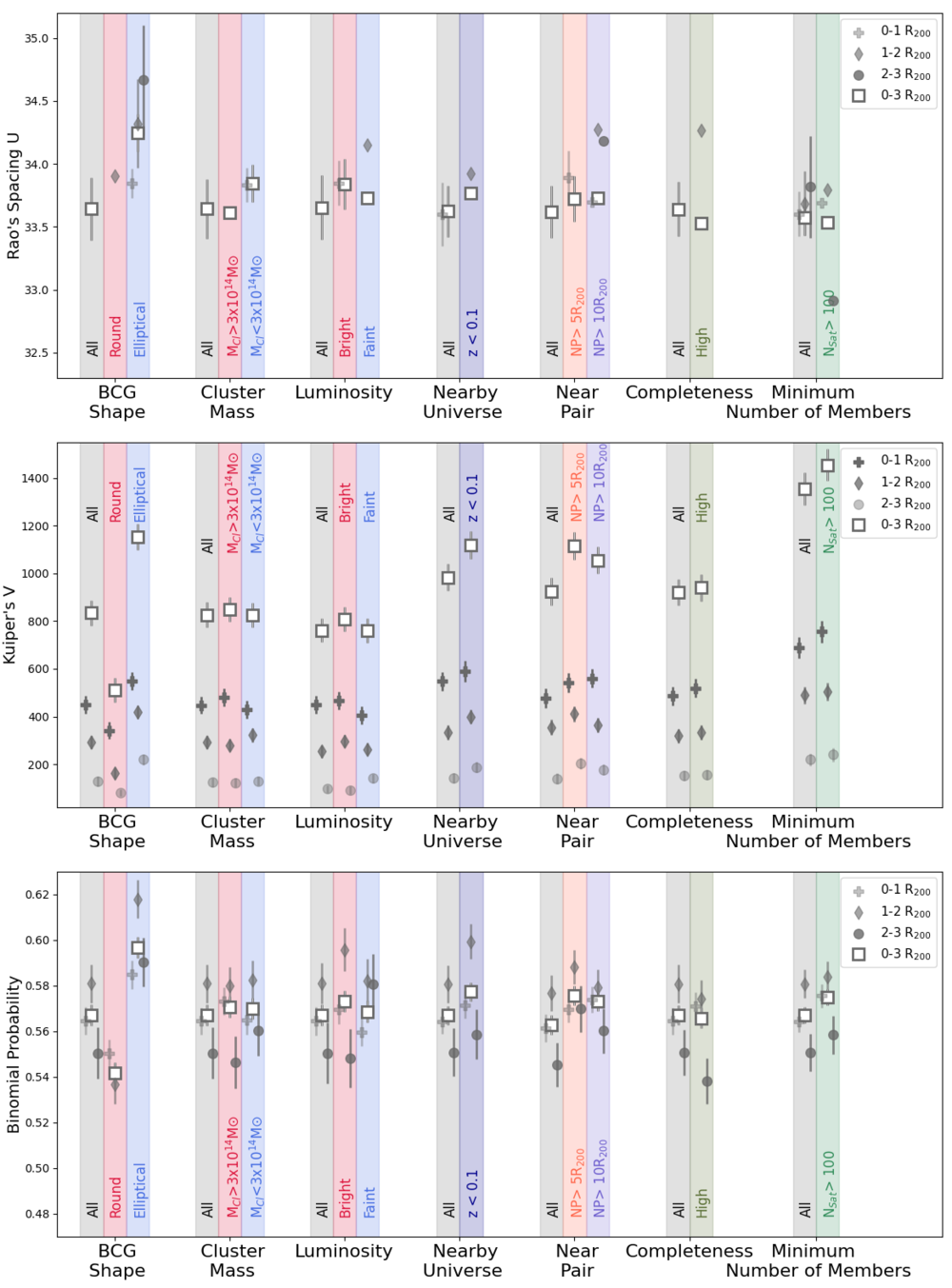}%S2Ncomp_satsonly.png}
\end{center}
\caption{Comparing the alignment signal strength for the `Cluster Members-only' sample. The y-axis is the value of $U$, $V$ and $P$ measure from top panel to bottom. The data symbols indicate the radial bin (see legend; $\rm 0-1~R_{200}$, $\rm 1-2~R_{200}$, $\rm 2-3~R_{200}$ and $\rm 0-3~R_{200}$). Subsamples that can be compared for equal sample size are grouped together along the x-axis. See text in Section \ref{sec:membersonlyresults} for a description of each parameter. Please note that sample sizes vary between radial bins, and between subsamples (e.g., between `BCG shape' and `Cluster Mass'). Therefore, comparisons should only be done within a subsample and for the same radial bin.}
\label{fig:UVPsatsonly}
\end{figure*}

\item {\it{Nearby Universe: }} In `Nearby Universe' (fourth column) we take a low redshift subsample labelled `z $<~0.1$', containing 108 of the 211 clusters. For the combined radial bins (0-3 R$_{200}$), the low redshift sample shows consistently higher alignment signal for all three measures ($U$, $V$ and $P$) although it is of low significance for the $U$ measure. For example, comparing clusters with no redshift limit versus clusters with redshift $<~0.1$, we measure $U=33.62$ versus $U=33.77$ (typical errors of $\sim0.20$), $V=982.6 \pm 56.3$ versus $V=1118.7 \pm 57.1$, and $P=0.567 \pm 0.004$ versus $P=0.577 \pm 0.004$ respectively (where all measurements have $P_{\rm{uniform}}=$ 1:million or less except the $U$ measures which have $P_{\rm{uniform}}=$ 1:60 and 1:3700 respectively). It is unclear if this represents a true demonstration that the alignment signal is evolving with the age of the Universe. The difference in maximum lookback-time between the  `All' and `Nearby Universe' samples is only $\sim$2~Gyr, therefore any evolutionary change would have to be very recent. We test the possibility that there might be larger numbers of satellites with measured redshifts in nearby clusters. But, we find little evidence for this or for differences in completeness when we compare the `All' and `Nearby Universe' sample, which gives additional weight to the hypothesis that we may in fact be measuring some true (recent) time evolution in the alignment signal.

\item {\it{Near Pair: }} Here, we take subsamples based on how close is the nearest cluster (labelled `Near Pair'). As with the LSS segments in Section \ref{sec:LSSsubsamples}, for the combined radial bins, there is some evidence for stronger alignment signal when some kind of restriction on the distance to the nearest cluster is included. Although, for the `Satellites only' sample, this result is only significant for the $V$ and $P$ measures. For example, comparing clusters with no limit on nearby clusters versus clusters where the nearest cluster must be more than 5 R$_{200}$ away, we measure $U=33.62 \pm 0.21$ versus $U=33.72 \pm 0.18$, $V=925.1 \pm 58.0$ versus $V=1116.3 \pm 58.9$, and $P=0.563 \pm 0.004$ versus $P=0.576 \pm 0.004$ respectively (where all measurements have $P_{\rm{uniform}}=$ 1:million or less except the $U$ measures which have $P_{\rm{uniform}}=$ 1:5). The preference for stronger alignment when the separation must be at least 5 R$_{200}$ (labelled `NP$>$5R$_{200}$') is less significant than was seen in the `Cluster members + LSS' sample.

\item {\it{Cluster Completeness: }} Here, we subsample clusters whose total completeness is high (second column from the right, labelled `Completeness'). We first measure the average completeness of galaxies within a 6-by-6 R$_{200}$ square, centred on the cluster (i.e., out to a projected radius of $\sim3$~R$_{200}$). Those with an average completeness $>$ 0.75 are classified as `High' completeness, and this subsample contains 104 of the 211 clusters. In general, we do not see a strong dependency of alignment signal on this parameter, for any of the alignment measures and across all the different radial bins, perhaps in part due to our efforts to correct the alignment signal for incompleteness (as described in Section \ref{sec:samples}). 

\item {\it{Minimum Number of Members: }} Finally, in the last column, we take a subsample where the number of cluster members must be $\ge$ 100. This subsample contains 112 of the 211 clusters. There is a small hint that this restriction may slightly increase the alignment signal for the combined radial bins of the $V$ and $P$ measures only but it is low significance.

\item \textit{Dependence on Projected Distance from Cluster: }
Similarly to our analysis of the LSS segments in Section \ref{sec:LSSsubsamples}, we now match the radial bin sample sizes to see how the signal strength varies with radius, independent of the effects of changing sample size, for the `All' model only. In contrast to the LSS segments (whose signal decreased with projected radius for both the $V$ and $P$ measures), we do not see a clear trend with projected radius for any of the three measurements of alignment strength. The results for $U$ are not significant. But, for $V$ and $P$ the results are significant and the strongest alignment signal arises in the second radial bin of 1-2~R$_{\rm{200}}$. For the radial bins (0-1, 1-2, 2-3)R$_{200}$, $V$ = (320.7, 382.6, 274.6) with typical errors of $\sim$30.0, and $P$ = (0.564, 0.581, 0.550) with typical errors of $\sim$0.007. This may indicate that the general trend for decreasing alignment signal with radius is reversed near the cluster center, as a result of a greater degree of virialization of those cluster members. Alternatively, galaxies near the cluster center are more likely to have fallen in earlier, when the alignment signal was perhaps less strong.

\end{itemize}

In summary, for the strength of the alignment signal between the BCG PA and the cluster member locations on the sky out to 3~R$_{200}$, the most significant parameter we find is the BCG shape. We see this result across all three measures of alignment signal ($U$, $V$ and $P$), and often in multiple radial bins. This strong dependence on BCG shape qualitatively agrees with the results for BCG alignment with the LSS segments in Section \ref{sec:LSSsubsamples}. We also see evidence for increased alignment signal in our low redshift sample across the three measures but with less significance than the dependency on BCG shape. Finally, there is some evidence for an increase in alignment signal strength when clusters with another cluster within 5~R$_{200}$ are removed from the sample in the case of the measures $V$ and $P$.

\subsection{Searching for alignment between the isophotal PAs of cluster members and their BCG}

In \cite{Huang18}, it was reported  that a signal of alignment between a cluster BCG PA and the satellite galaxies PAs exists, although other studies do not find a clear indication of this beyond the second or third brightest galaxy in the cluster \citep{Torlina2007,NiedersteOstholt10, Sifon15,West17}. We note that this differs from the alignment we have been measuring. In their study they measure the isophotal PAs (rather than the positional PAs on the sky) of the individual cluster members. They found that this type of alignment was stronger for brighter satellites and those closer to the cluster centre. The redshift range of their sample is roughly comparable with ours and consists of relatively nearby clusters ($z<$ 0.35).

\begin{figure}%[ht]
\includegraphics[width=93mm]{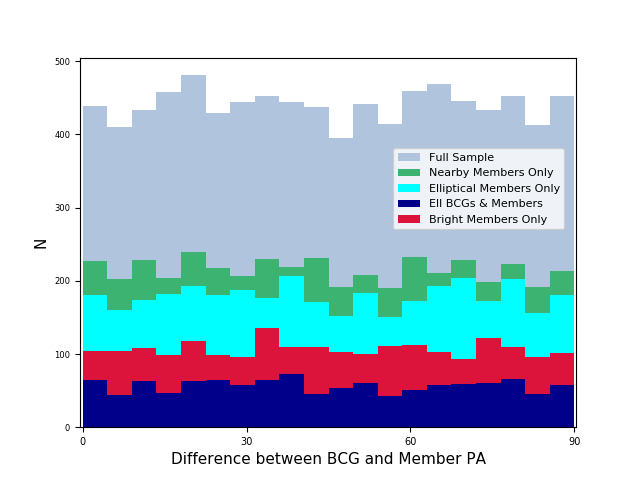}%BCGsatalign_histo.png}
\caption{Distribution of the angle between the satellite and BCG PA for the `Cluster Members-only' sample (`Full Sample' in the legend). We also show the following subsamples (indicated in the legend); `Nearby Members Only' for cluster members that are within 1~R$_{200}$ of the cluster centre, `Elliptical Members Only' and `Ell BCGs \& Members' including only galaxies with r-band axial ratio $b/a<0.75$, and `Bright members only' including only cluster members with M$_{\rm{r}}<-20.5$.}
\label{fig:BCSvssatPA}
\end{figure}

Inspired by their initiative, we attempted to make the same measurement using our `Cluster Members-only' sample. In Figure \ref{fig:BCSvssatPA}, we plot histograms of the difference in the PA angle between the BCG and their cluster members isophotal PAs. If there is no alignment, and cluster members have completely random isophotal PAs that are independent of their BCG PA, then we expect a flat distribution. To maximise the statistics, we use our `Cluster Members-only' sample (labelled `Full Sample' in the legend). We see no clear evidence for the cluster members to show any preference for particular isophotal PAs with relation to their BCG PA. Our three statistical measurements of alignment (Rao's spacing test, Kuiper's V test, and the Binomial Probability Test) all find the distribution of cluster member isophotal PAs is highly consistent with a uniform distribution.

We also try dividing up the sample into various subsamples. The `Nearby Members Only' subsample is cluster members within 1~R$_{200}$ only. We also consider a sample where the cluster members must have an elliptical shape so as the PA should be clearly defined, and where both the cluster members and BCGs must be elliptical (labelled `Elliptical Members Only' and `Ell BCGs \& Members', respectively). As defined previously, an elliptical shape means an axial ratio $b/a<$ 0.75. Finally, we consider a subsample with only bright cluster members (M$_{\rm{r}}<-20.5$; labelled `Bright Members only'). In all cases, the cluster member BCGs have flat distributions, as if they are randomly orientated with respect to their BCG. Measurements of the $U$, $V$ and $P$ values give probabilities of being uniform that are too high to pass our criteria for significance as described in Section \ref{sec:significance} (e.g., For the $P$ measure, $P_{\rm{uniform}}>0.93$ for the `Full sample', meaning it is highly probable to be consistent with a uniform distribution). 

Although we see no clear evidence for cluster-satellite isophotal PA alignment in our data set, we note that our statistics are poorer than in the \cite{Huang18} study. We have only 211 clusters compared to their several thousands of clusters, and our satellite sample is roughly one-tenth the size of their sample. Nevertheless, our spectroscopy is deeper, and we only used spectroscopically confirmed cluster members, meaning our membership criteria is more strict and accurate. Thus, our sample could potentially provide an interesting data set for this experiment. It is interesting that we cannot detect this type of BCG-satellite isophotal PA alignment, and yet we detect such significant signals of BCG alignment with the positions of the cluster members and the LSS about the cluster within the same sample. We conclude that this is a clear demonstration of how much stronger the alignment signal is with positions of the cluster members and the LSS compared to the isophotal PA alignment between BCG and satellites.

\section{Discussion and Summary}
Using a sample of 211 clusters whose cluster members have been determined using deep spectroscopy, we search for a well-known signal of alignment between the position angle (PA) of the Brightest Cluster Galaxy (BCG) and the locations of the cluster members on the sky about the cluster centre (referred to as `BCG-cluster' alignment). The deep spectroscopy provides us with large numbers of cluster members in each cluster (typically $>$ 100 members) which should make our sample ideal to detect alignment signals. Furthermore, using the overlap with Sloan Digital Sky Survey (SDSS) imaging, we can make corrections for incompleteness to reduce the possibility that the measured alignment is not artificially a result of uneven completeness levels about the cluster centers.   

We combine this data set with a 3D map of the skeleton of the Large Scale Structure (LSS) derived using SDSS data. This allows us to search for a direct link between the BCG PA and the location of the LSS segments on the sky at distances far beyond the cluster members. By using the skeleton of the LSS, we effectively filter out noise from scatter in individual galaxy positions in the distribution of the LSS. We also remove filaments that are not interconnected with the main filaments that directly connect to the cluster of interest. As a result, we should be more sensitive to any signal of alignment and we use this data set to search for alignment signal out to projected radii as large as $\rm 6-10~R_{200}$.

BCG alignment is measured using three individual statistical measures; Rao's spacing test, Kuiper's V-statistic, and the Binomial probability (denoted as the measures $U$, $V$ and $P$ respectively). We test the non-uniformity of a measurement by computing the probability that the measure is reproduced by randomly sampling PAs from a uniform distribution with a sample of equal size to the observed sample. 

We also divide our sample up into subsamples according to a wide range of parameters, to try to measure the dependency of the alignment strength on those parameters. When comparing the strength of the alignment signal between subsamples, we match the size of the subsamples to avoid variations in signal strength due to differing sample size.

Our key results are given in the following:
\begin{enumerate}

\item The BCG alignment signal for a stack of all the LSS segments available is very high significance (1:$>$million chance of being drawn from a uniform distribution) for the $U$, $V$ and $P$ measure. This is true in all of the individual bins of projected radius as well, including our most distant at 6-10~R$_{200}$ from the cluster centre\footnote{with the sole exception of the innermost radial bin of the $U$ measure which has 1:$>$100 thousand chance of being uniform}.

\item The shape of the cluster BCG is a key parameter determining the strength of the alignment signal. BCGs that appear more elliptical show significantly stronger alignment with both their cluster members and also with the surrounding LSS segments. When all radial bins are combined, this is visible for all three measures of the alignment signal ($U$, $V$ and $P$), and is generally visible in the individual projected radial bins as well, even at 6-10~R$_{200}$ from the cluster centre.  
%\newline 

\item The BCG alignment signal with the LSS segments increases significantly when we require that a filament connects with the cluster at 1~R$_{200}$ instead of 3~R$_{200}$, and also when we use a luminosity-weighted LSS skeleton instead of simply the number density of galaxies. The alignment signal appears stronger for two of our measures ($V$ and $P$) when we remove clusters from the sample with a nearby companion cluster, and this holds true for BCG alignment with both the cluster members and LSS segments. There is also a hint that clusters connected with three or more filaments have slightly stronger alignment signal but it is a low-significance result.

\item For the BCG alignment with the members of the cluster, the alignment signal increases for two of our measures ($V$ and $P$) when we consider a low redshift subsample (z$<0.1$) although the significance is not as high as seen for other parameters mentioned above. Meanwhile, we see no clear dependency (or only a weak, low-significance dependency) on parameters such as the cluster mass or galaxy luminosity.

%From the other parameters we considered, such as cluster member luminosity and cluster completeness, we do not see a clear impact on S/N$_{\rm{equal}}$. Similar to \cite{Huang16}, we find no indication of a dependency on redshift, although the range of our sample redshift is rather limited ($z=0.003-0.289$) compared to some studies that claim to detect evolution (e.g., \citealt{NiedersteOstholt10, Hao2011}). 

%Of possible value to future studies using LSS skeletons to study alignment, we note that S/N$_{\rm{equal}}$ appears to increase slightly (generally by $\sim$ 20\% or less) when we use a luminosity-weighted LSS skeleton instead of a galaxy number density-weighted one. The S/N$_{\rm{equal}}$ also increases by a similar amount when we force the skeleton to meet within 1~R$_{200}$ of the cluster. However, this latter condition comes at the expense of cluster sample size, and so was not applied in our fiducial sample.

\end{enumerate}

The fact that the BCG tends to be preferentially aligned with LSS at projected distances as large as $\rm 6-10~R_{200}$ is telling. So called `backsplash' galaxies (galaxies that have previously entered a sphere of radius 1~R$_{200}$ from the cluster but are now found at larger radius; \citealt{Gill2005}) typically don't reach further than $\sim \rm 2-3~R_{200}$ from the cluster centre. Therefore, we physically interpret our results as evidence that the preferential feeding of galaxies into clusters along connected filaments must build-up the cluster BCG. And, this process must continue throughout the time period when a significant fraction of the BCG stars were put in place. If the merger-axis is preferentially aligned with the filaments, the stellar body of the BCG may become extended in that same direction, and growth could potentially occur through a combination of minor and major mergers. Similarly, the galaxies that are fed into the clusters but don't merge with the BCG will tend to form a population of cluster members that are similarly extended along the direction of the filaments. We note that the continuous feeding of new material into the cluster, along the preferential directions of filaments, would be expected to eventually change the shape of the main cluster dark matter halo itself, if the new mass is a significant fraction of the total.

Finally, we also search for evidence that the PAs of the cluster members (based on their shape) might be aligned with their BCG PA (e.g, \citealt{Huang18}). However, we find no clear evidence for this in our sample, and our results are consistent with the PAs of the cluster member being randomly orientated with respect to the PA of their BCG. This demonstrates how much weaker this type of alignment is compared to the BCG-cluster and BCG-LSS alignment.

If mergers along preferential directions are indeed responsible for the clear alignment we see between the PA of the BCG and the locations of cluster members and the LSS skeleton, it is interesting to ask if we might expect a dependency of the alignment signal strength on the cluster dynamical state. We plan to present the results of such an analysis in a following paper, where we will separate the cluster sample according to their dynamical state and quantify the strength of the alignment with the cluster members and surrounding LSS.

\section*{Acknowledgements}
We thank the referee for a careful reading of the draft and constructive comments that improved the paper. We thank St\'ephane Rouberol for the smooth running of the HORIZON Cluster, where some of the post-processing was carried out.
We thank Thierry Sousbie for provision of the DisPerSE code (\href{http://ascl.net/1302.015}{ascl.net/1302.015}). R.S. acknowledges support from Fondecyt Regular Grant project number 1230441. J.Y. was supported by a KIAS Individual Grant (QP089901) via the Quantum Universe Center at Korea Institute for Advanced Study. J.W.K. acknowledges support from the National Research Foundation of Korea (NRF), grant No. NRF-2019R1C1C1002796, funded by the Korean government (MSIT). H.S.H. was supported by the National Research Foundation of Korea (NRF) grant funded by the Korea government (MSIT) (No. 2021R1A2C1094577). K.K. acknowledges support from the DEEPDIP project (ANR-19-CE31-0023). 
Funding for the Sloan Digital Sky 
Survey IV has been provided by the 
Alfred P. Sloan Foundation, the U.S. 
Department of Energy Office of 
Science, and the Participating 
Institutions. 
SDSS-IV acknowledges support and 
resources from the Center for High 
Performance Computing  at the 
University of Utah. The SDSS 
website is www.sdss.org.
SDSS-IV is managed by the 
Astrophysical Research Consortium 
for the Participating Institutions 
of the SDSS Collaboration including 
the Brazilian Participation Group, 
the Carnegie Institution for Science, 
Carnegie Mellon University, Center for 
Astrophysics | Harvard \& 
Smithsonian, the Chilean Participation 
Group, the French Participation Group, 
Instituto de Astrof\'isica de 
Canarias, The Johns Hopkins 
University, Kavli Institute for the 
Physics and Mathematics of the 
Universe (IPMU) / University of 
Tokyo, the Korean Participation Group, 
Lawrence Berkeley National Laboratory, 
Leibniz Institut f\"ur Astrophysik 
Potsdam (AIP),  Max-Planck-Institut 
f\"ur Astronomie (MPIA Heidelberg), 
Max-Planck-Institut f\"ur 
Astrophysik (MPA Garching), 
Max-Planck-Institut f\"ur 
Extraterrestrische Physik (MPE), 
National Astronomical Observatories of 
China, New Mexico State University, 
New York University, University of 
Notre Dame, Observat\'ario 
Nacional / MCTI, The Ohio State 
University, Pennsylvania State 
University, Shanghai 
Astronomical Observatory, United 
Kingdom Participation Group, 
Universidad Nacional Aut\'onoma 
de M\'exico, University of Arizona, 
University of Colorado Boulder, 
University of Oxford, University of 
Portsmouth, University of Utah, 
University of Virginia, University 
of Washington, University of 
Wisconsin, Vanderbilt University, 
and Yale University.

%%%%%%%%%%%%%%%%%%%%%%%%%%%%%%%%%%%%%%%%%%%%%%%%%%
\section*{Data Availability}
The data underlying this article will be shared on reasonable request to the corresponding author.



\bibliographystyle{mnras}
\bibliography{bibfile} % if your bibtex file is called example.bib

%\begin{thebibliography}{99}
%\end{thebibliography}

%%%%%%%%%%%%%%%%%%%%%%%%%%%%%%%%%%%%%%%%%%%%%%%%%%
%%%%%%%%%%%%%%%%% APPENDICES %%%%%%%%%%%%%%%%%%%%%

\appendix

\section{Additional completeness tests}
\label{sec:completeness}
\begin{figure}%[ht]
\begin{center}
\includegraphics[width=90mm]{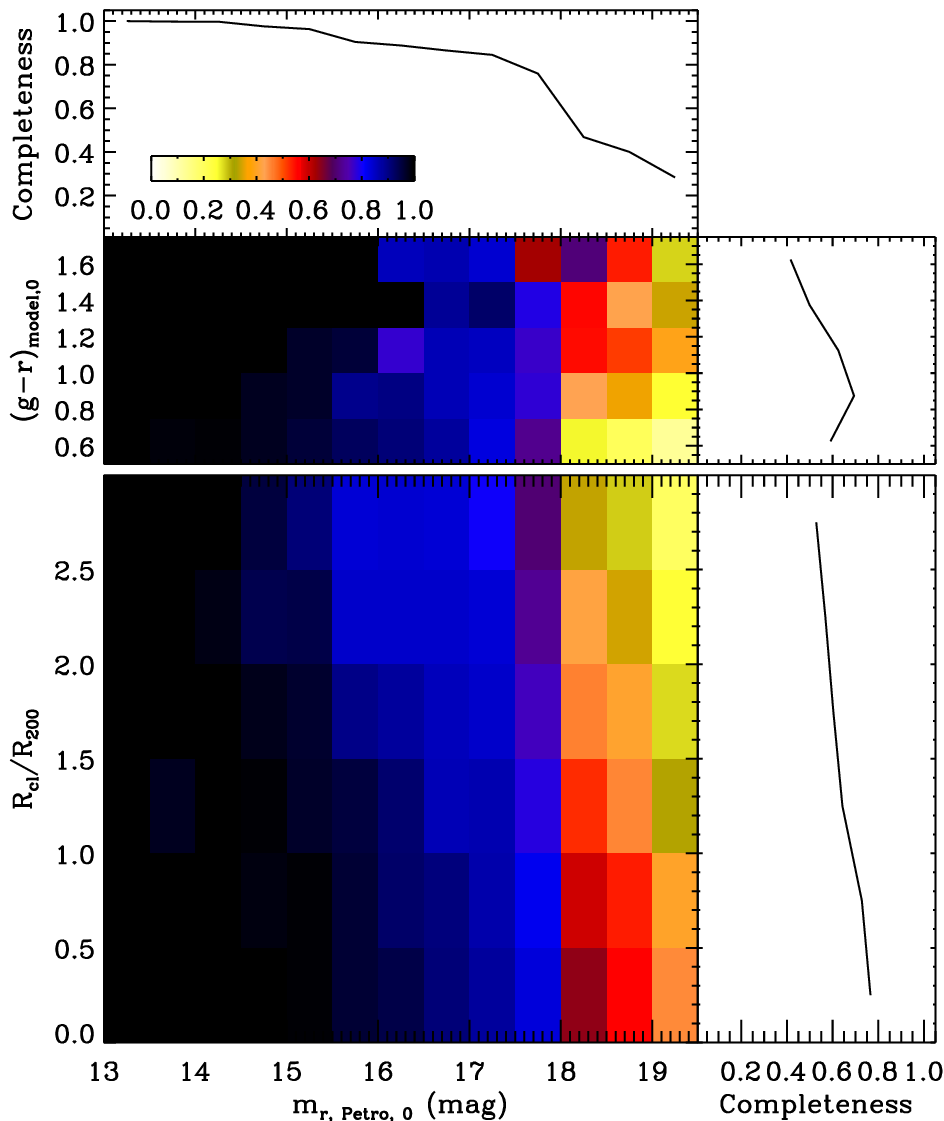}
\caption{Spectroscopic completeness as a function of a galaxy's Petrosian magnitude in the r-band (x-axis), (g-r) colour (y-axis, upper panel) and clustercentric radius normalised by R$_{200}$ (y-axis, lower panel) for the full sample of clusters. From top to bottom, the side panels show  curves of completeness versus r-band magnitude, versus colour, and versus normalised clustocentric radius. For the (g-r) colour, `model' magnitudes are used (instead of Petrosian magnitudes) to ensure equivalent apertures in both filters.}
\label{fig:Completeness_newfig}
\end{center}
\end{figure}

In Figure \ref{fig:Completeness_newfig}, we plot the completeness (shown in the colour-bar) of all the galaxies in the `Cluster Members only' sample. On the x-axis they are plotted as a function of r-band petrosian magnitude. In the bottom panel, the y-axis shows the projected distance from the cluster centre normalised by R$_{200}$. The completeness can be seen to decrease with projected distance. For example, at a magnitude of r=15, completeness falls from $\sim1.0$ to $\sim0.8$ at 3 R$_{pr}/$R$_{200}$ (see also the side panel where we marginalise over the galaxy magnitude). Nevertheless, by measuring the completeness in a grid with cell side lengths of 1~$R_{200}$ surrounding the cluster, and then correcting for incompleteness, we have already normalised out this radial dependency in our analysis.

In any case, the existence of a radial gradient should not influence our main conclusions unless it would preferentially affected galaxies in a particular direction with respect to the position angle of the cluster BCG. We see no evidence that the completeness varies as a function of angle from the BCG position angle in our two main samples. An example of this is shown in the top-right panel of Figure \ref{fig:stackedall3Rvirconnect}. In the bottom subpanel, we plot the completeness as a function of angle from the BCG position angle. There is a different line for each radial bin, but they are all nearly horizontal when plotted against the BCG position angle. There is also no evidence that they show higher completeness nearer the BCG position angle, something that would be necessary to artificially induce an alignment signal. We also plot histograms of the number of galaxies in angular bins about the BCG position angle for the `Cluster Members + LSS' sample (different panels are different radial bins). The solid lines are before the completeness correction and the filled histograms are after the completeness correction, and both the histograms show a similar alignment signal. Thus, we expect that our results would not differ strongly, even if we had not conducted our completeness correction.

In the top panel of Figure \ref{fig:Completeness_newfig}, the (g-r) galaxy colour is shown on the y-axis. The side panel shows that the completeness peaks at (g-r)$\sim$0.9. This is to be expected as the cluster galaxies selected from the various HeCS surveys were primarily red sequence selected \citep{RedSeqTech} based on SDSS panchromatic photometry. Naturally this means the sample is biased towards early-type galaxies. Several previous studies have demonstrated that the signal of alignment is stronger in red galaxies \citep[][etc]{Rykoff14,Georgiou2019}, perhaps because galaxies become preferentially redder towards the centres of cosmological filaments \citep{{Kraljic2018},Winkel2021}. Therefore, we would expect that our sample would be even more sensitive to the alignment signal between cluster members and the BCG position angle. However, we note that we used galaxy redshifts from the SDSS survey to build the skeletons of the LSS. These were not red sequence selected, therefore, unlike the cluster members, the alignment signal between the BCG position angle and positions of filament segments on the sky should not be colour biased in the same manner. Thus, caution is required when directly comparing the alignment signal strength between the cluster members versus the LSS, as noted in Section \ref{sec:LSSsubsamples}.

As a final note on completeness, we also measured an overall completeness for each cluster individually. This is the average completeness of galaxies within a 6-by-6 R$_{200}$ square, centered on the cluster. Those with an average completeness $>$0.75 were subsampled and classified as `high' completeness. In Figure \ref{fig:UVPsatsonly}, we test how the relative strength of the alignment signal between the `All' sample and `high' completeness sample varies. We see no clear dependency on the cluster completeness for any of our three measures of the alignment signal, or in any of the subsamples by radial bin range that we consider.

\section{Table with full Sample of Clusters considered in this study}
\label{sec:tableofclusters}
A complete list of the clusters considered in this study is provided in Table \ref{tab:clusterlist}. The `Cluster Members only' sample combines all these clusters. The `Cluster Members + LSS' sample uses only clusters labelled as `LSS' in column (xvi) of the Table. A description of each individual column of the Table follows. Further details can be found in Section \ref{samplesection}. Column (i) gives the cluster name. Column (ii)-(iii) are the right ascension and declination of each cluster, as provided by their respective survey. The Table is ordered by increasing right ascension.  Column (iv) is the mean redshift of the cluster members, as identified using the caustic technique \citep{Diaferio1999}. Columns (v)-(vii) are cluster mass, velocity dispersion, and R$_{200}$. Column (viii) shows the r-band apparent magnitude limit that was used for the cluster satellites. Column (ix) is the number of spectroscopically confirmed cluster members. Column (x) is the spectroscopic survey from which the cluster members were selected. Columns (xi-xv) are the properties of the cluster BCG, right ascension, declination, ellipticity, position-angle and r-band absolute magnitude. Finally, column (xvi) indicates if the cluster is located in the main SDSS area where we have built skeletons of the surrounding Large Scale Structure (LSS).
%\begin{adjustwidth}{-2.5 cm}{-2.5 cm}\centering\begin{threeparttable}[!htb]...\end{threeparttable}\end{adjustwidth}

% \usepackage{lscape}
% \usepackage{longtable}
% Note: It may be necessary to compile the document several times to get a multi-page table to line up properly

% \usepackage{lscape}
% \usepackage{longtable}
% Note: It may be necessary to compile the document several times to get a multi-page table to line up properly

% \usepackage{lscape}
% \usepackage{longtable}
% Note: It may be necessary to compile the document several times to get a multi-page table to line up properly

\clearpage
\onecolumn

\begin{landscape}
\setlength{\tabcolsep}{4pt} % Default value: 6pt
%\renewcommand{\arraystretch}{1.0} % Default value: 1
%
\end{landscape}

%\clearpage
%\onecolumn

% \usepackage{lscape}
% \usepackage{longtable}
% Note: It may be necessary to compile the document several times to get a multi-page table to line up properly
%\newpage

\clearpage
\section{Tables with number of objects in each subsample}
\label{sec:Nsamp}

%\FloatBarrier
\begin{table*}
\centering
\caption{Number of segments of the LSS in each subsample of the `Cluster Members + LSS' sample from Figure \ref{fig:UVPsatsandLSS}}
\label{tab:samplesize_LSS}
\begin{tabular}{c|ccccc}
\backslashbox{Radius}{Subsample} & BCG shape & Nconnect & R$_{200}$,connect & Near Pair & LSS weight \\
\hline
$[0 - 3]$ R$_{200}$    & 958      & 235      & 2033        & 1112       & 1728        \\
$[3 - 6]$ R$_{200}$     & 808      & 303      & 1702        & 1044       & 1480        \\
$[6 - 10]$ R$_{200}$    & 1591     & 515      & 3018        & 1650       & 2334        \\
$[0 - 10]$ R$_{200}$    & 3357     & 1053     & 6753        & 3806       & 5542       
\end{tabular}
\end{table*}
%\FloatBarrier

%\FloatBarrier
\begin{table*}
\centering
\caption{Number of galaxies in each subsample of the `Cluster Members only' sample from Figure \ref{fig:UVPsatsonly}}
\label{tab:samplesize_gals}
\begin{tabular}{c|ccccccc}
\backslashbox{Radius}{Subsample} & BCG shape & Cluster Mass & Luminosity & Nearby Universe & Near Pair & Completeness & N$_{\textrm{mem,min}}$ \\
\hline
$[0 - 1]$ R$_{200}$ & 6728  & 6252  & 6262  & 7679  & 7175  & 6800  & 9683  \\
$[1 - 2]$ R$_{200}$& 3373  & 3369  & 2909  & 3871  & 4229  & 3699  & 5739  \\
$[2 - 3]$ R$_{200}$ & 1988  & 1905  & 1449  & 2197  & 2554  & 2371  & 3526  \\
$[0 - 3]$ R$_{200}$ & 11639 & 11526 & 10620 & 13747 & 13958 & 12872 & 18948
\end{tabular}
\end{table*}
%\FloatBarrier

%\FloatBarrier
\begin{table*}
\centering
\caption{Number of clusters in each subsample of the `Cluster Members + LSS' sample in Figure \ref{fig:UVPsatsandLSS}}
\label{tab:Ncluster_LSS}
\begin{tabular}{cccccccccc}
Fiducial & \multicolumn{2}{c}{BCG shape} & \multicolumn{3}{c}{Nconnect} & R200,connect & \multicolumn{2}{c}{Near Pair}  & LSS Weighting \\
         & Round         & Ellip         & =2   & =3   & $\ge$4 & 1            & \textgreater{}10 R$_{200}$ & No Limit & No Weight      \\
\hline         
91       & 43            & 48            & 24  & 28  & 39               & 64           & 55                   & 124     & 86       \\    
\end{tabular}
\end{table*}
%\FloatBarrier

%\FloatBarrier
\begin{table*}
\centering
\caption{Number of clusters in each subsample of the `Cluster Members only' sample in Figure \ref{fig:UVPsatsonly}}
\label{tab:Ncluster_gals}
\begin{tabular}{cccccccccccc}
All &
  \multicolumn{2}{c}{BCG shape} &
  \multicolumn{2}{c}{Cluster Mass} &
  \multicolumn{2}{c}{Luminosity} &
  Nearby Universe &
  \multicolumn{2}{c}{Near Pair} &
  Completeness &
  N$_{\textrm{mem,min}}$ \\
 &
  Round &
  Ellip &
  High &
  Low &
  Bright &
  Faint &
  z\textless{}0.1 &
  $>$10~R$_{200}$ &
  $>$5~R$_{200}$ &
  High &
  N$_{sat}$\textgreater{}100 \\
\hline
211 &
  105 &
  106 &
  74 &
  137 &
  211 &
  211 &
  108 &
  131 &
  172 &
  104 &
  112
\end{tabular}
\end{table*}
%\FloatBarrier


%%%%%%%%%%%%%%%%%%%%%%%%%%%%%%%%%%%%%%%%%%%%%%%%%%

% Don't change these lines
\bsp	% typesetting comment
\label{lastpage}
\end{document}